\documentclass[%
 pre,
 twocolumn,
 superscriptaddress,
 amsmath,amssymb,
 aps,
 floatfix
]{revtex4-2}

\usepackage{graphicx}
\usepackage{dcolumn}
\usepackage{bm}
\usepackage[colorlinks=true, allcolors=blue]{hyperref}

\begin{document}

\preprint{APS/123-QED}

\title{Arrhenius temperature dependence of the crystallization time of deeply supercooled liquids}

\author{Yuki Takaha}
 \email{yukitakaha@g.ecc.u-tokyo.ac.jp}
 \affiliation{Graduate School of Arts and Sciences, The University of Tokyo, 3-8-1 Komaba, Tokyo 153-8902, Japan}
\author{Hideyuki Mizuno}
 \affiliation{Graduate School of Arts and Sciences, The University of Tokyo, 3-8-1 Komaba, Tokyo 153-8902, Japan}
\author{Atsushi Ikeda}
 \affiliation{Graduate School of Arts and Sciences, The University of Tokyo, 3-8-1 Komaba, Tokyo 153-8902, Japan}
 \affiliation{Research Center for Complex Systems Biology, Universal Biology Institute, The University of Tokyo, Komaba, Tokyo 153-8902, Japan}

\date{\today}

\begin{abstract}
Usually, supercooled liquids and glasses are thermodynamically unstable against crystallization. 
Classical nucleation theory (CNT) has been used to describe the crystallization dynamics of supercooled liquids. 
However, recent studies on overcompressed hard spheres show that their crystallization dynamics are intermittent and mediated by avalanche-like rearrangements of particles, which largely differ from the CNT. 
These observations suggest that the crystallization times of deeply supercooled liquids or glasses cannot be described by the CNT, but this point has not yet been studied in detail. 
In this paper, we use molecular dynamics simulations to study the crystallization dynamics of soft spheres just after an instantaneous quench. 
We show that although the equilibrium relaxation time increases in a super-Arrhenius manner with decreasing temperature, the crystallization time shows an Arrhenius temperature dependence at very low temperatures. 
This is contrary to the conventional formula based on the CNT. 
Furthermore, the estimated energy barrier for the crystallization is surprisingly small compared to that for the equilibrium dynamics.
By comparing the crystallization and aging dynamics quantitatively, we show that a coupling between aging and crystallization is the key for understanding the rapid crystallization of deeply supercooled liquids or glasses. 
\end{abstract}

\maketitle

\section{Introduction}
Classical nucleation theory (CNT) effectively describes first-order phase transition dynamics.
This theory predicts that nucleation first takes place, where sufficiently large droplets of the ordered phase are created in the sea of the disordered phase and then the droplets grow steadily~\cite{debenedetti1997metastable,onuki2002phase,Binder_1987}. 
The nucleation is driven by thermal fluctuations and the nucleation time $\tau_{\mathrm{N}}$, the inverse of the nucleation rate, is given by 
\begin{eqnarray}
\tau_{\mathrm{N}} &\sim& \tau_t \exp\left(\Delta G/k_{\mathrm{B}}T\right), \label{eq:nucleation}
\end{eqnarray}
where $k_{\mathrm{B}}$ is Boltzmann's constant and $T$ is the temperature. 
The constant $\tau_t$ is the characteristic time scale of the transport process and $\Delta G$ is the free energy barrier to form the critical droplet, which is given by the maximum value of the droplet formation energy as a function of the droplet size. 
In its simplest version, the droplet formation energy is given by the combination of the free energy gain due to the low energy ordered phase and the surface tension between two phases.   
The CNT has been successfully applied to a wide variety of systems undergoing first-order phase transitions. 
In the Ising model, for example, the direct numerical estimate of $\tau_{\mathrm{N}}$ was shown to accurately be described by Eq.~(\ref{eq:nucleation})~\cite{Stauffer_1982,Shneidman_1999,Ryu_2010}.

For the crystallization of slightly supercooled liquids, the formula Eq.~(\ref{eq:nucleation}) is known to work well~\cite{lundrigan2009test,chkonia2009evaluating}, although the first-principle theoretical prediction of $\tau_t$  and $\Delta G$ is difficult and still under debate~\cite{Auer_2001,Kawasaki_2010}. 
As liquids are further supercooled, the glass transition is approached, which results in a dramatic slowing of the particle transport process. 
To effectively take this into account in the CNT, $\tau_t$ is conventionally approximated by the equilibrium relaxation time $\tau_{\alpha}$~\cite{cavagna2009supercooled}~\footnote{
Although the inverse of the diffusion constant $D^{-1}$ is a better approximation of $\tau_t$~\cite{Tanaka_2003,Kawasaki_2010}, we omit the difference between $D^{-1}$ and $\tau_{\alpha}$. 
These two quantities show slightly different temperature dependences due to the Stokes-Einstein violation~\cite{cavagna2009supercooled}, but their difference is still minor for our later discussion.}, 
leading to
\begin{eqnarray}
\tau_{\mathrm{N}} &\sim& \tau_{\alpha} \exp\left(\Delta G/k_{\mathrm{B}}T\right). \label{eq:convention}
\end{eqnarray}
Notably, this formula predicts that $\tau_{\mathrm{N}}$ is a nonmonotonic function of temperature. 
The nucleation time $\tau_{\mathrm{N}}$ diverges as the temperature increases since the first-order transition is approached and the free energy gain diminishes so that $\Delta G$ diverges.
On the other hand, $\tau_{\mathrm{N}}$ drastically increases as the temperature decreases since the glass transition is approached and $\tau_{\alpha}$ drastically increases. 
As a result, $\tau_{\mathrm{N}}$ has a minimum value at a certain temperature. 
Reflecting this, the CNT predicts a similar nonmonotonic temperature dependence for the crystallization time $\tau_{\mathrm{cry}}$, which is the time required for crystalline regions to occupy a major part of the sample. 
Indeed, several experimental~\cite{masuhr1999time,kim1996experimental,schroers2001transition} and numerical~\cite{levchenko2011molecular,moore2011structural} studies confirmed that the nucleation and crystallization time has a minimum value at a certain temperature.

Despite these successes of the CNT, it has recently been reported that the crystallization of deeply supercooled liquids and glasses is qualitatively different from that in the CNT.
In highly overcompressed hard spheres, crystallization proceeds before the transport of particles becomes diffusive~\cite{zaccarelli2009crystallization}. 
The crystallization is caused by small displacements of particles, which are even smaller than the particles' diameter~\cite{sanz2011crystallization}. 
Moreover, avalanche-like crystallization has been reported for well annealed ``mature'' glasses in hard spheres~\cite{sanz2014avalanches} and pseudo hard spheres~\cite{de2017brownian,rosales2016avalanche}, in which intermittent stochastic rearrangements of particles cause crystallization. 
Some similarities between avalanche-like crystallization dynamics and aging dynamics have been reported~\cite{yanagishima2017common}. 
An experiment of a colloidal system also shows observations consistent with these results~\cite{ganapathi2021structure}.
These phenomena do not appear in the CNT, which assumes equilibrium conditions and a continuum picture. 

These peculiar crystallization dynamics are interesting, but quantitative analysis of the crystallization time is still lacking. 
In particular, the temperature dependence of the crystallization time has not yet been studied in detail.
This is partly because the previous studies focused mainly on hard spheres, where the temperature does not play any role. 
In this paper, we perform molecular dynamics (MD) simulations of soft spheres. 
We mainly focus on the dynamics of crystallization after an instantaneous quench from a high temperature to low temperatures. 
We show that, while the equilibrium relaxation time depends on the temperature in a super-Arrhenius manner, the crystallization time shows an Arrhenius-like temperature dependence at very low temperatures. 
Moreover, the corresponding energy barrier is much smaller than that for the equilibrium dynamics. 
This clearly contradicts Eq.~(\ref{eq:convention}). 
We show that the crystallization process at low temperatures consists of an early process, in which the system falls into the nearest inherent structure, and a late process, in which transitions between inherent structures occur. 
The crystallization time is dominated by the late process, which is widely different from the continuum picture in the CNT. 
Furthermore, we show that although this late process has some similarities with the aging dynamics, the crystallization time cannot be directly explained by the characteristic time scales of the aging dynamics. 
Instead, a coupling between the aging and crystallization dynamics is the key for understanding the rapid crystallization.

\section{Methods}

We perform MD simulations of monodisperse and polydisperse soft spheres. 
The interaction potential is an inverse-power-law (IPL) potential given by 
\begin{eqnarray}
v(r_{ij})&=&\varepsilon\Bigg\lbrack\left(\frac{\sigma_{ij}}{r_{ij}}\right)^{12}-\left(\frac{\sigma_{ij}}{r_{c,ij}}\right)^{12}\nonumber\\
 && +A\left(\frac{r_{ij}-r_{c,ij}}{\sigma_{ij}}\right)+B\left(\frac{r_{ij}-r_{c,ij}}{\sigma_{ij}}\right)^2\Bigg\rbrack,
\end{eqnarray}
where $\varepsilon$ is an energy scale; $r_{ij}=\left|\bm{r}_i-\bm{r}_j\right|$, with $\bm{r}_i$ being the position of particle $i$; and $\sigma_{ij}\equiv\left(\sigma_i+\sigma_j\right)/2$, where $\sigma_{i}$ represents the effective diameter of particle $i$.
The coefficients $A$ and $B$ force the continuity of the first and second derivatives at the cutoff length $r_{c,ij}\equiv1.5\sigma_{ij}$.
For polydispersity, we introduce a top-hat distribution of particle size with the average particle size $\bar{\sigma}$ and the width $\Delta$, in which a non-trivial particle arrest is known not to occur~\cite{zaccarelli2015polydispersity}. 
We consider the systems with $\Delta=0.00,~0.12,~0.24,$ and $0.36$ in this work, where $\Delta=0.00$ corresponds to the monodisperse system. 
To approximate a bulk system, we use the periodic boundary condition in a cubic box (volume $V$).
Mass, length, time, and temperature are measured in units of $m$, $\bar{\sigma}$, $\bar{\sigma}\sqrt{m/\varepsilon}$ and $\varepsilon/k_{\mathrm{B}}$, respectively, where $m$ is the particle mass and $k_{\mathrm{B}}$ is the Boltzmann constant. 
Throughout this paper, the temperature is controlled by the Nos\'{e}-Hoover thermostat, where the thermostat mass is tuned so that unphysical temperature oscillation decays sufficiently rapidly even after the instantaneous quench. 
The time step of the MD simulation is 0.01.
The number of particles $N$ is $16384$ unless otherwise noted.
The packing fraction $\phi=\left(\sum_{i}\pi\sigma_i^3/6)\right/V$ is fixed to be $\pi/6$ for both monodisperse and polydisperse systems. 
At this density, the freezing temperature is $T_{\mathrm{Freezing}}=0.59$ for the monodisperse system~\footnote{We calculated the free energy of the equilibrium liquid and crystal with Monte Carlo simulations. The calculations were conducted according to the method described in Ref.~\cite{prestipino2005phase}, which we do not describe in detail.}.

To measure the degree of crystallization of the samples, we use the method introduced in Ref.~\cite{zaccarelli2009crystallization,pusey2009hard}.
First, each particle is assigned a vectorial bond-order parameter $d_6$~\cite{wolde1996simulation}. 
Pairs of neighbors are then identified by using a cutoff distance $1.4 \sigma_{ij}$, and a bond of a pair is deemed ``solid-like" if the scalar product of their $d_6$ vectors exceeds 0.7.
A particle is labeled solid-like if it has at least six solid connections.
Finally, we calculate the crystallinity $X(t)$ of a sample as the proportion of solid-like particles at a given time $t$.

\section{Results and discussion}
\subsection{Equilibrium dynamics}

\begin{figure}
\centering
\includegraphics[width=0.95\columnwidth]{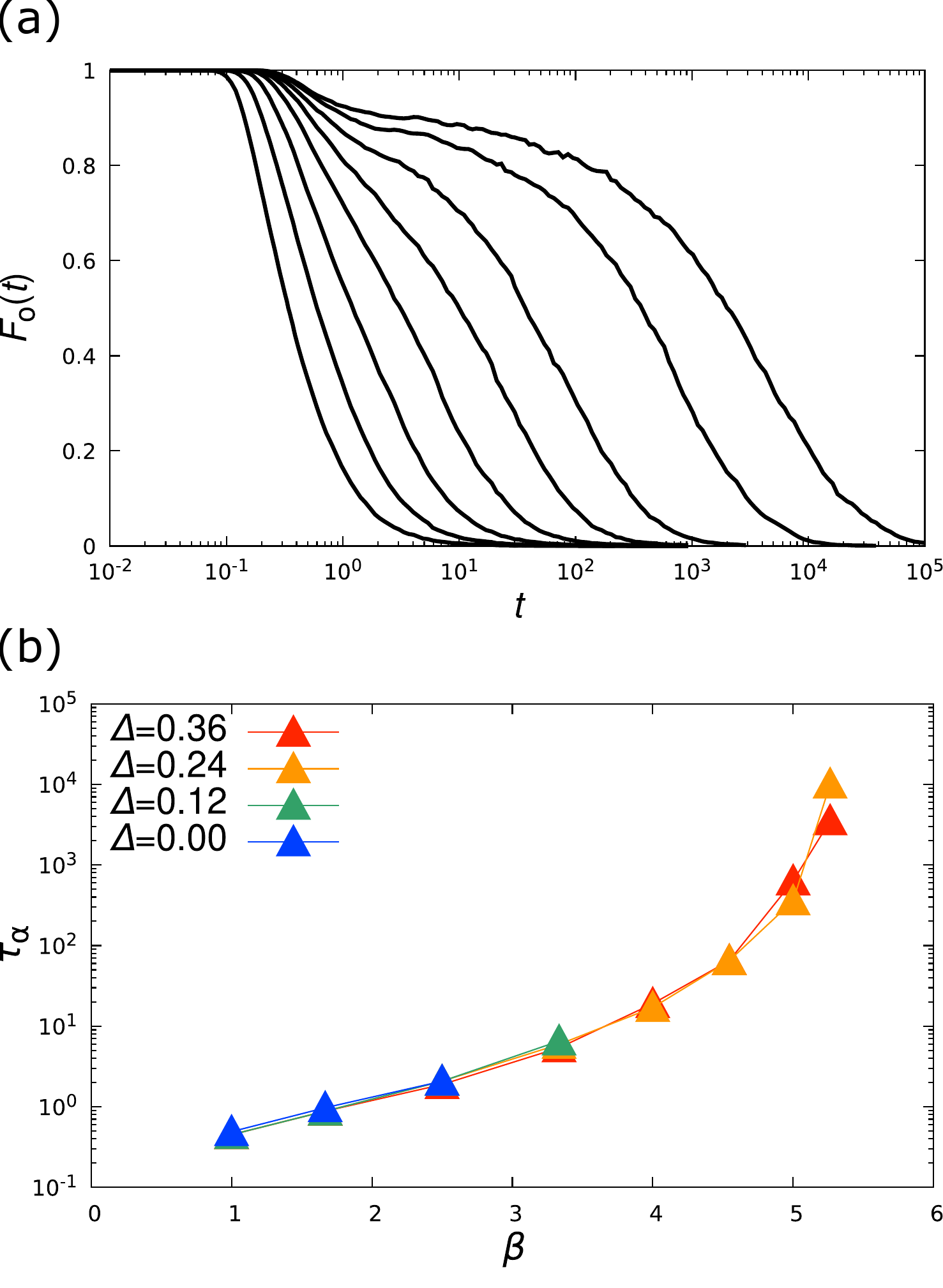}
\caption{(a) Overlap functions with $\Delta=0.36$.
The temperatures of the correlation functions correspond to $T=1.00,~0.60,~0.40,~0.30,~0.25~,0.22,~0.20$ and $0.19$ from left to right.
(b) Relaxation time of equilibrium dynamics of the liquids with $\Delta=0.00,~0.12,~0.24$ and $0.36$.
The inverse temperature $\beta\equiv 1/T$ is used on the horizontal axis.}
\label{fig:eq}
\end{figure}

To discuss the crystallization dynamics on firm ground, we first clarify the equilibrium dynamics of the model. 
We equilibrate the system at a target temperature $T$ and polydispersity $\Delta$ and then perform the production runs starting from the equilibrium configurations. 
We perform 20 independent runs at each state point.
The overlap function $F_{\mathrm{O}}(t)=\langle\sum_n\Theta(0.3-|\bm{r}_n(t)-\bm{r}_n(0)|)/N\rangle$ is then calculated from the time series of the particle configurations, where the bracket $\langle\rangle$ represents the average over 20 samples. 
This calculation is performed with $N=2048$. 

Figure~\ref{fig:eq}(a) shows the overlap functions for $\Delta=0.36$. 
Clearly, the overlap function decays rapidly at higher $T$ and the relaxation becomes much slower with decreasing $T$.
It shows a two-step relaxation, which is a hallmark of the dynamics of supercooled liquid.
We measure the equilibrium relaxation time $\tau_\alpha$ defined as $F_{\mathrm{O}}(t=\tau_\alpha)=0.4$ for various state points ($T$, $\Delta$). 
For some $(T,~\Delta)$ conditions, we find that the system undergoes crystallization during the equilibration or production runs. 
Since we focus solely on the equilibrium dynamics here, we do not calculate $\tau_\alpha$ if $\langle X(t=\tau_\alpha)\rangle\ge0.5$. 
We note that the equilibration time is always 40 times larger than the relaxation time $\tau_{\alpha}$.

Figure~\ref{fig:eq}(b) shows the temperature dependence of the relaxation time $\tau_{\alpha}$ for various $\Delta$. 
The relaxation time $\tau_{\alpha}$ clearly shows a super-Arrhenius temperature dependence, although the relaxation time cannot be calculated with small $\Delta$ and low $T$ because of crystallization.
Importantly, the relaxation times for various $\Delta$ collapse onto almost the same curve, meaning that $\tau_{\alpha}$ does not strongly depend on $\Delta$. 
In previous studies, it has been reported that the fragility depends on the polydispersity; the more polydisperse the system, the stronger the dynamics~\cite{kawasaki2007correlation}.
We can indeed find such a tendency, albeit slightly, when we closely compare $\Delta=0.24$ and $0.36$.
However, this effect is minor, as we focus on the small $\Delta$ region, and we can regard the relaxation time as almost independent from $\Delta$.

\subsection{Crystallization time}
\begin{figure}
\centering
\includegraphics[width=0.95\columnwidth]{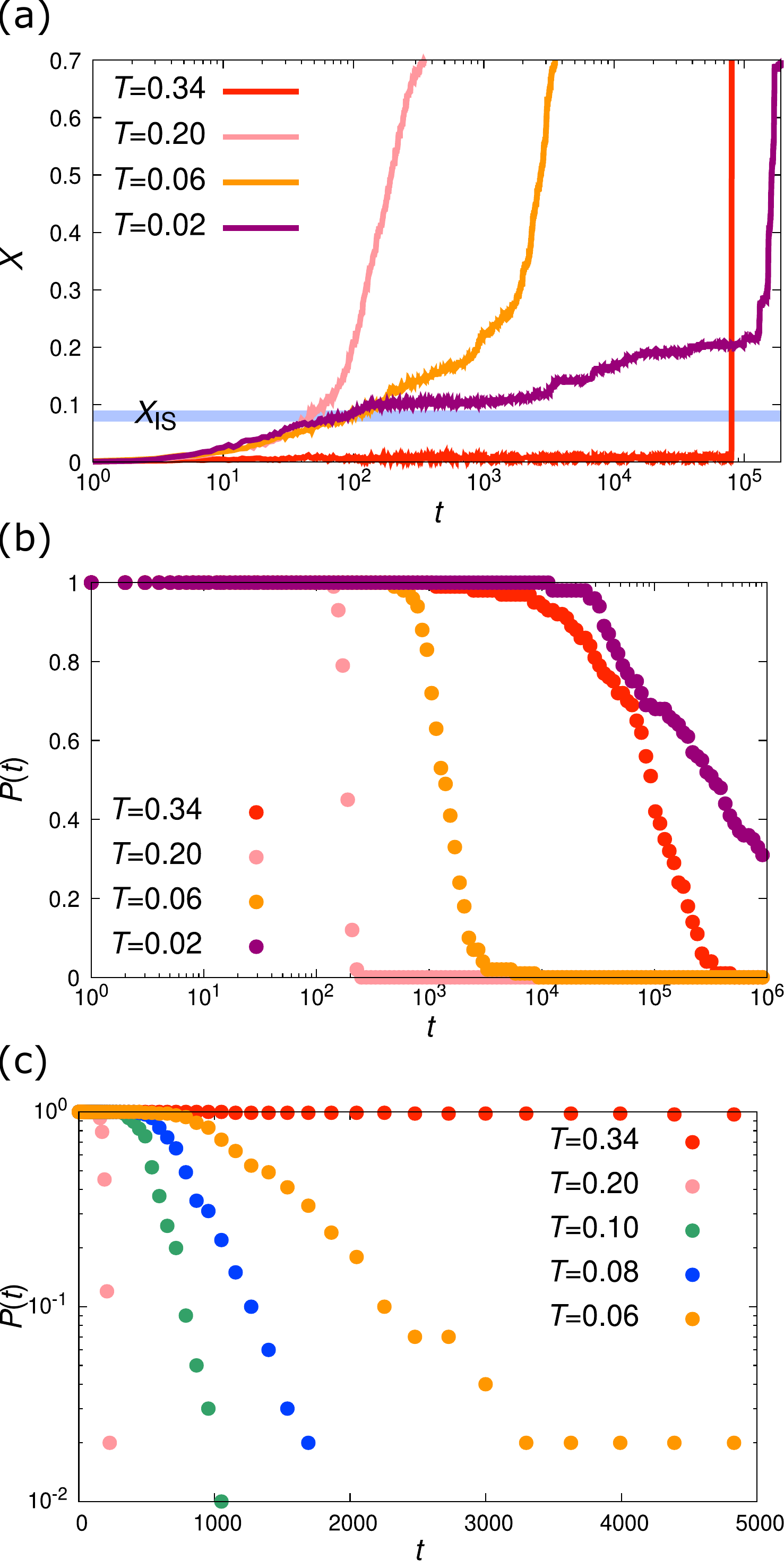}
\caption{(a) Time series of crystallinity $X(t)$.
The blue shadow represents the average $X_{\mathrm{IS}}$ of crystallinity of the inherent structures of the equilibrium liquid configurations with $T=3.00$.
(b)(c) Survival probability $P(t)$, which is the probability of uncrystallized samples.}
\label{fig:crydynamics}
\end{figure}

We now focus on the crystallization dynamics of the monodisperse system.
We first prepare equilibrium liquid configurations at $T=3$ and then perform MD simulations at the target temperatures starting from these configurations. 
This corresponds to an instantaneous quench from $T=3$ to the target temperatures. 
We show the results with $N=16384$, which is free from a finite size effect~\footnote{
Note that the crystallization dynamics become slower in much smaller systems at lower temperatures. 
This slowing is a finite size effect due to the collisions of the crystalline regions which the periodic boundary condition facilitates.
We confirmed that $N=16384$ is large enough for the focused temperature range. }. 
Figure~\ref{fig:crydynamics}(a) shows the time evolution of the crystallinity $X(t)$ of typical runs. 
At $T=0.34$, the crystallinity $X(t)$ fluctuates around $X=0$ for a long time, and then it suddenly starts to increase at very large $t$. 
This time evolution is qualitatively consistent with the dynamics predicted by the CNT, which consist of rare nucleation and steady growth of the nucleus.
On the other hand, $X(t)$ for $T=0.20,~0.06,~0.02$ is very different. 
The crystallinity $X(t)$ shows an increase up to $10$\% in the short-time region $t \lesssim 10^2$, which we call the early process of the crystallization. 
Interestingly, $X(t)$ for various temperatures coincides in this time region, meaning that the early process does not strongly depend on the temperature if the temperature is sufficiently low. 
At larger time $t \gtrsim10^2$, gradual and slow increases are observed, which we call the late process of the crystallization. 
Unlike the early process, the late process becomes slower and slower with decreasing temperature. 
At the lowest temperature, $X(t)$ fluctuates around $10$\% and increases intermittently.
Then, it finally grows at very large $t$. 
These results mean that the intermittent crystallization takes place not only in overcompressed hard spheres~\cite{sanz2014avalanches} but also in soft spheres at low temperature. 
These crystallization dynamics clearly differ from that described by the CNT.

To discuss the statistical law that governs the crystallization, we calculate the survival probability~\cite{chkonia2009evaluating}. 
We repeat the simulations described above for 100 independent samples and calculate the crystallinity $X(t)$ for each of them. 
We label the samples as ``crystallized" at a given time $t$ if the crystallinity $X(t)$ becomes larger than $0.5$~\footnote{
We confirmed that an increase or decrease in the threshold value 0.5 only causes a constant-fold change in the crystallization time, barely altering its temperature dependence.}.
Then, we calculate the survival probability $P(t)$, which is defined as the proportion of noncrystallized samples at time $t$.
Figures~\ref{fig:crydynamics}(b) and \ref{fig:crydynamics}(c) show the survival probability $P(t)$ at each temperature.
Figure~\ref{fig:crydynamics}(c) shows that $P(t)$ does not decrease from 1 in a certain time, called a lag time, and then decays exponentially at larger $t$. 
This means that crystallization of a given sample can be seen as a random event with a constant rate, although there is a minimum latency time. 
Note that similar behavior of $P(t)$ was previously reported for the liquid-gas condensation dynamics in the Lennard-Jones system, where the CNT works well~\cite{chkonia2009evaluating}. 
In this case, the lag time was identified as the time required for the growth of the nucleus, and the exponential decay time was identified as the nucleation time.  

\begin{figure}
\centering
\includegraphics[width=0.95\columnwidth]{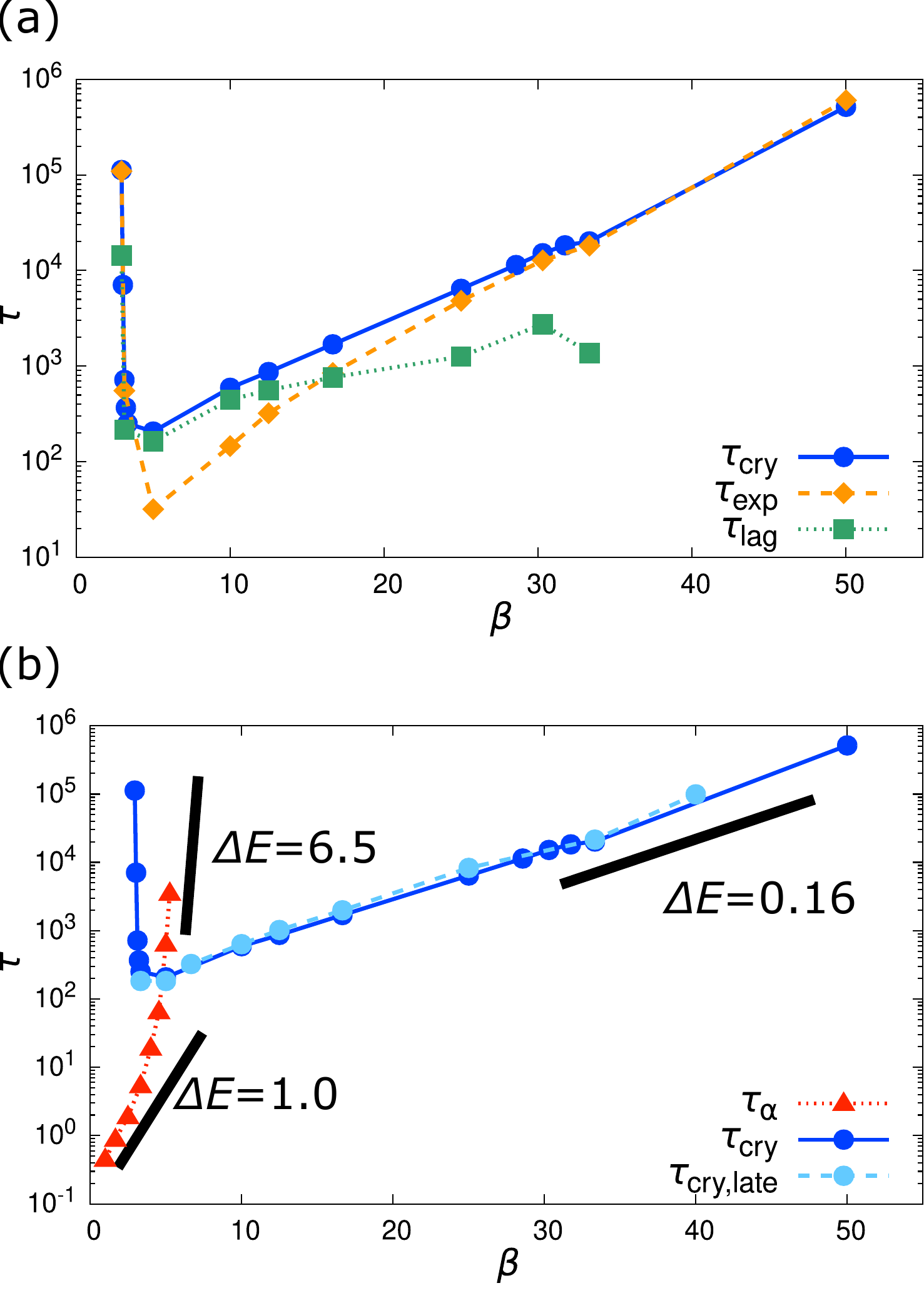}
\caption{(a) Temperature dependence of $\tau_{\mathrm{cry}},\ \tau_{\mathrm{exp}}$, and $\tau_{\mathrm{lag}}$.
(b) Temperature dependence of the crystallization time and equilibrium relaxation time.
The inverse temperature $\beta\equiv 1/T$ is used on the horizontal axis.
The time scale $\tau_{\mathrm{cry,late}}$ represents the crystallization time when the initial configurations were the ISs of the liquids with $T=3.0$.
The three black lines represent the slope of $\exp(\Delta E/T)$.}
\label{fig:taucry}
\end{figure}

We now extract the characteristic time scales of the crystallization. 
The simplest definition of the crystallization time is the time at which the survival probability reaches some threshold value. 
In practice, we define $\tau_{\rm cry}$ as $P(t=\tau_{\mathrm{cry}})= 0.4$, the time at which 60 \% of samples become crystallized. 
Additionally, we can define the lag time $\tau_{\rm lag}$ and the exponential decay time $\tau_{\rm exp}$ by fitting $P(t)$ to $\exp((t - \tau_{\rm lag})/\tau_{\rm exp})$~\footnote{
This fitting was done for the data in $0.3 < P(t) < 0.8$.
Note that the error of the estimate of $\tau_{\mathrm{lag}}$ became larger when $\tau_{\mathrm{exp}}$ became much larger than $\tau_{\mathrm{lag}}$, and we could not estimate $\tau_{\mathrm{lag}}$ at the lowest temperature.
}.
These three time scales are plotted in Fig.~\ref{fig:taucry}(a). 
At approximately the freezing temperature, the exponential decay time $\tau_{\rm exp}$ is much larger than the lag time $\tau_{\rm lag}$ and then $\tau_{\rm cry} \approx \tau_{\rm exp}$. 
This means that rare nucleation is the rate-controlling process and growth is much quicker. 
With decreasing temperature, $\tau_{\rm exp}$ drops more sharply than $\tau_{\rm lag}$ and then $\tau_{\rm lag}$ dominates $\tau_{\rm cry}$. 
In this temperature range, nucleation becomes fast and comparable to the microscopic time scale, which may be called ``spinodal nucleation''~\cite{pusey2009hard}. 
Finally, at the lowest temperatures, the exponential decay time $\tau_{\rm exp}$ increases rapidly, while the lag time $\tau_{\rm lag}$ does not, and as a result, $\tau_{\rm cry}$ is again dominated by $\tau_{\rm exp}$.  
Therefore, our results establish that the exponential decay time $\tau_{\rm exp}$ is the unique time scale of crystallization in the low-temperature regime where the intermittent crystallization takes place, and this time scale can be well captured by $\tau_{\rm cry}$. 
The irrelevance of the lag time suggests that any deterministic process, such as the crystal growth, is absent or at least is not a rate-controlling process in this temperature regime.

Finally, we quantitatively discuss the temperature dependence of the crystallization time. 
For comparison, the equilibrium relaxation time of polydisperse system $\Delta=0.36$ is also plotted in Fig.~\ref{fig:taucry}(b). 
From the viewpoint of Eq.~(\ref{eq:convention}), $\tau_{\rm cry}$ of the monodisperse system should be compared with $\tau_{\alpha}$ of the monodisperse system, but this cannot be done because $\tau_{\alpha}$ cannot be measured for the monodisperse system at low temperatures due to rapid crystallization. 
However as shown in Sec.~III, $\tau_{\alpha}$ does not strongly depend on the polydispersity in our model, so $\tau_{\alpha}$ of the monodisperse system can be well approximated by those of polydisperse systems~\cite{zaccarelli2009crystallization}.
The crystallization time $\tau_{\mathrm{cry}}$ diverges not only at the freezing point $T_{\mathrm{freezing}}=0.59$ but also at lower temperatures and thus has a minimum at a finite temperature (see Fig.~\ref{fig:taucry}(b)).
This is qualitatively consistent with the conventional crystallization theory.
However, at sufficiently low temperature, the crystallization time becomes much smaller than the equilibrium relaxation time, $\tau_{\alpha}\gg\tau_{\mathrm{cry}}$. 
Importantly, the temperature dependence of the crystallization time is Arrhenius-like, while that of the equilibrium relaxation time increases in a super-Arrhenius manner. 
This is contrary to Eq.~(\ref{eq:convention}). 
By fitting the crystallization time and the equilibrium relaxation time with $\tau\sim\exp(\Delta E/T)$, we estimate the characteristic energy barrier $\Delta E$ for these processes. 
The energy barrier $\Delta E$ for the crystallization is $0.16$ while $\Delta E$ for the equilibrium relaxation of the high-temperature and low-temperature liquids are $1.0$ and $6.5$, respectively (see Fig.~\ref{fig:taucry}(b)). 
The energy barrier $\Delta E$ for the crystallization is surprisingly small, which is even smaller than that for the equilibrium dynamics of high-temperature liquids. 
Therefore, we conclude that crystallization in deeply supercooled liquids proceeds by crossing energy barriers that are much smaller than those experienced by the equilibrium liquids.

\subsection{Early process and late process}
The early and late processes, which we have divided the crystallization dynamics into above, can be qualitatively distinguished from the viewpoint of the potential energy landscape.
We apply the optimization method FIRE~\cite{bitzek2006structural} to the equilibrium configurations at $T=3$ and obtain their inherent structures (ISs). 
For 20 independent samples, we measure the crystallinity $X$ of these ISs and define their average as $X_{\mathrm{IS}}$.
While the crystallinity $X$ of the original equilibrium liquid configurations is almost zero, $X_{\mathrm{IS}}$ is approximately 8\%.
This value is almost equal to the crystallinity reached by the early process (see Fig.~\ref{fig:crydynamics}(a)).
This suggests that the early process corresponds to the dynamics in which the system falls into the nearest IS. 
This is consistent with the observation that the dynamics in the early process hardly depend on the temperature. 

The observation above also suggests that the late process of crystallization consists of transitions between the ISs. 
To directly access this dynamics, we perform simulations that mimic the late process; namely, we perform MD simulations at the target temperatures starting from the ISs. 
Here, the initial momenta are generated following the Maxwellian distribution at the target temperatures. 
We perform 20 independent runs, monitor the time evolution of $X(t)$ and analyze the survival probabilities as described in the previous section. 
By this, we obtain the crystallization time for this dynamics, which we call $\tau_{\mathrm{cry, late}}$. 
As shown in Fig.~\ref{fig:taucry}(b), this crystallization time $\tau_{\mathrm{cry,late}}$ is equal to the original crystallization time $\tau_{\mathrm{cry}}$.
Therefore, we conclude that the early process (falling into ISs) is not relevant for the crystallization time and that the late process (transitions between ISs) dominates the crystallization time in deeply supercooled liquids.

\subsection{Aging and crystallization}
The inequality $\tau_{\alpha}\gg\tau_{\mathrm{cry}}$ implies that the deeply supercooled liquid crystallizes before equilibration.  
This naturally explains the reason why Eq.~(\ref{eq:convention}) does not work because it assumes that the transport of particles is controlled by the equilibrium dynamics. 
This observation suggests that the crystallization time would be better compared with the relaxation time of the system just after an instantaneous quench, which is the aging dynamics. 
To discuss this possibility, we perform the following analysis. 
For monodisperse $\Delta=0.00$ and polydisperse $\Delta=0.36$ systems, we first prepare the equilibrium configurations at $T=3$, apply FIRE to obtain the corresponding ISs, and perform MD simulations at target temperatures starting from these ISs as described in the previous section. 
We then measure the overlap function $F_{\mathrm{O}}(t)$ to monitor the relaxation dynamics. 
This protocol is similar to the one used in the studies of the aging dynamics for zero waiting time, although crystallization is expected to take place during the simulation runs in the monodisperse case. 
Note that we perform simulations for 20 independent samples and the results are averaged over them. 

\begin{figure}
\centering
\includegraphics[width=0.95\columnwidth]{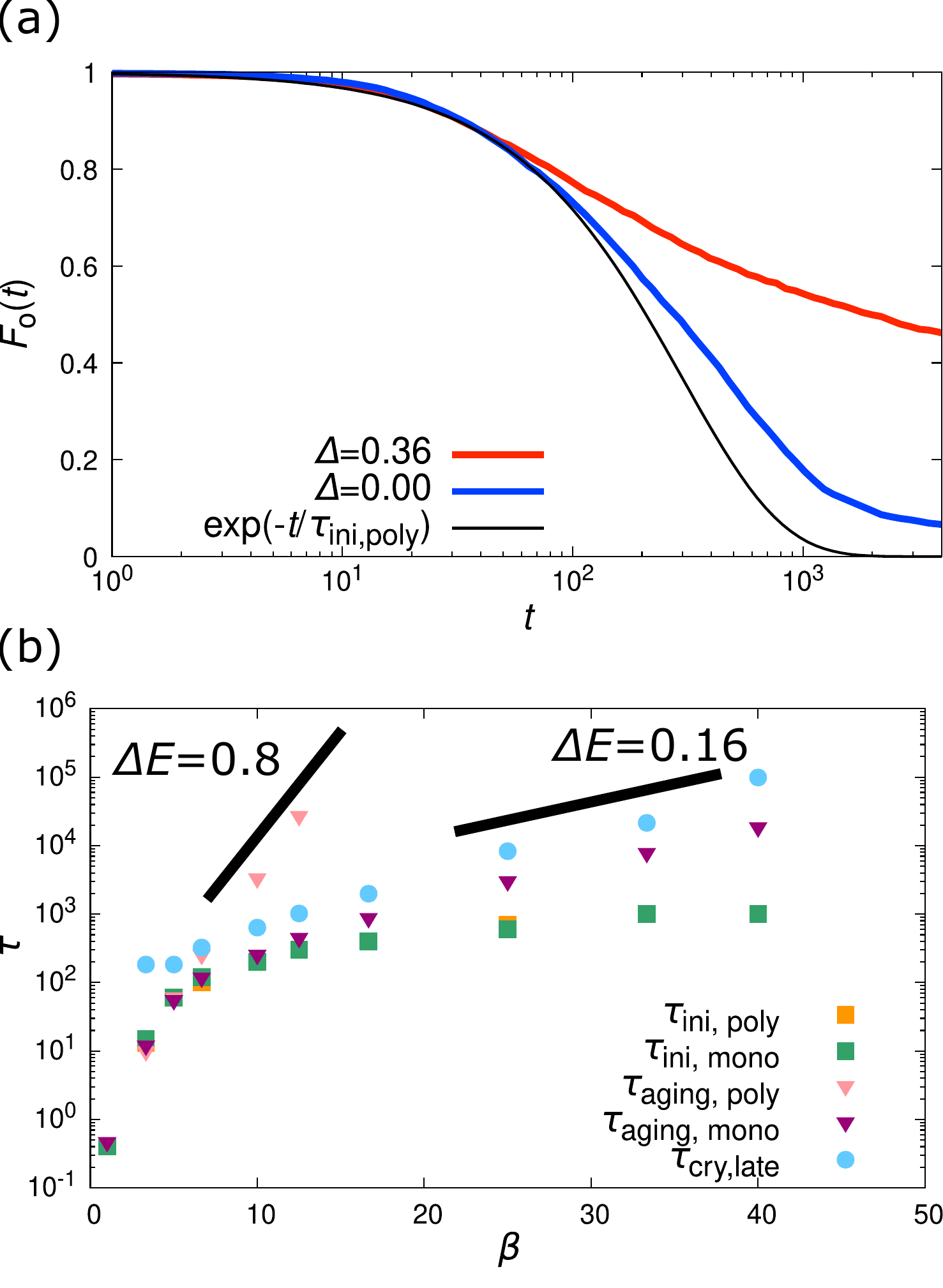}
\caption{(a) Overlap functions of nonequilibrium dynamics with $T=0.080$.
The black line shows the fitting curve of $F_{\mathrm{O}}(t)$ of a monodisperse system with $\exp(-t/\tau_{\mathrm{ini,mono}})$.
(b) Relaxation times of nonequilibrium dynamics and crystallization time.
The inverse temperature $\beta\equiv 1/T$ is used on the horizontal axis.
The two black lines represent the slope of $\exp(\Delta E/T)$.}
\label{fig:aging}
\end{figure}

Figure~\ref{fig:aging}(a) shows the overlap functions $F_{\mathrm{O}}(t)$ of the monodisperse and polydisperse systems with $T=0.08$.
We first focus on the polydisperse case where crystallization does not occur.  
The correlation function has two properties that are not observed for the equilibrium dynamics.
(1) The short-time relaxation ($t \lesssim 10^2$).   
The overlap function $F_{\mathrm{O}}(t)$ starts to decrease at approximately $t=10$. 
This short-time relaxation is not due to the vibrations of particles but the rearrangements of particles; we checked that the amplitude of the vibrations is too small to affect the overlap function since the target temperature is very low. 
To extract the time scale $\tau_{\mathrm{ini,poly}}$ characterizing these earliest rearrangements, we fit $F_{\mathrm{O}}(t)$ with $\exp\left(-t/\tau_{\mathrm{ini,poly}}\right)$ in the short-time region. 
It is clear that the fitting works well in the region $F_{\mathrm{O}}(t) \gtrsim 0.8$, namely for the rearrangement of approximately the first 20\% of particles. 
(2) The long-time relaxation ($t \gtrsim 10^2$). 
The overlap function $F_{\mathrm{O}}(t)$ becomes widely separated from the exponential function and has a tail in the long-time regime. 
This means that the relaxation of the majority of particles is qualitatively different from that of the first 20\% of particles and the former cannot be seen as the repetition of the latter. 
Note that a similar long-time tail in the aging dynamics was reported in Ref.~\cite{el2010subdiffusion}.
We introduce the relaxation time in this time regime as $F_{\mathrm{O}}(t=\tau_{\mathrm{aging,poly}})=0.4$. 

The two relaxation times obtained for the polydisperse system, $\tau_{\mathrm{ini,poly}}$ and $\tau_{\mathrm{aging,poly}}$, are plotted in Fig.~\ref{fig:aging}(b). 
Unlike the equilibrium relaxation time, the relaxation time in the long-time regime $\tau_{\mathrm{aging,poly}}$ shows an Arrhenius-like temperature dependence at low temperatures. 
However, they are still larger than the crystallization time. 
An Arrhenius fit of this relaxation time yields the energy barrier $\Delta E = 0.8$, which is far from that for crystallization. 
On the other hand, the relaxation time in the short-time regime $\tau_{\mathrm{ini,poly}}$ is much smaller than the crystallization time. 
We find that its temperature dependence is almost linear in $1/T$ rather than Arrhenius-like, suggesting that the corresponding energy barrier is vanishingly small. 
Therefore, we conclude that the relaxation times in the aging dynamics do not give a direct explanation of the crystallization time; namely, one cannot improve Eq.~(\ref{eq:convention}) by replacing $\tau_{\alpha}$ with $\tau_{\mathrm{ini,poly}}$ or $\tau_{\mathrm{aging,poly}}$. 

Finally, we focus on the monodisperse case. 
The overlap function is shown in Fig.~\ref{fig:aging}(a). 
We define the relaxation time in the short-time $\tau_{\mathrm{ini,mono}}$ and long-time regimes $\tau_{\mathrm{aging,mono}}$ in the same way as the polydisperse case and plot them in Fig.~\ref{fig:aging}(b)~\footnote{
For simplicity, we use the symbol $\tau_{\mathrm{aging,mono}}$, although not only aging but also crystallization take place in this case.}.  
For the short-time decay, the overlap function of the monodisperse system is almost the same as that in the polydisperse system. 
Also, the relaxation times $\tau_{\mathrm{ini,mono}}$ and $\tau_{\mathrm{ini,poly}}$ coincide at all the temperatures studied. 
Therefore, crystallization does not affect the first rearrangements of particles in glasses. 
However, this relaxation time is widely separated from the crystallization time. 
On the other hand, in the long-time regime, the overlap function of the monodisperse system decays much faster than that in the polydisperse system. 
This means that the crystallization itself accelerates the relaxation dynamics. 
Moreover, the relaxation time $\tau_{\mathrm{aging,mono}}$ has a similar temperature dependence as the crystallization time. 
Indeed, a fitting of $\tau_{\mathrm{aging,mono}}$ with $\exp(\Delta E/T)$ yields $\Delta E=0.16$, which coincides with that for $\tau_{\mathrm{cry}}$. 
Therefore, the crystallization itself accelerates the relaxation dynamics, and the resulting relaxation time gives the crystallization time. 
This suggests that a coupling between the relaxation and crystallization dynamics is the key for understanding the rapid crystallization of glasses. 
We also find that the crystallinity $X(t)$ begins to increase on this time scale $\tau_{\mathrm{aging,mono}}$, which is consistent with the discussion above. 

We additionally mention that the equivalence between $\tau_{\mathrm{aging,mono}}$ and $\tau_{\mathrm{cry}}$ is not trivial because the former is measured from the overlap function, which does not monitor the crystallinity or other orders. 
It monitors only the small displacement (up to 0.3) of particles. 
This observation is consistent with Ref.~\cite{zaccarelli2009crystallization}, where the crystallization due to small displacements was reported.

\section{Conclusion}
In this paper, we used MD simulations to study the dynamics of monodisperse and polydisperse soft spheres, with a particular focus on the crystallization dynamics just after the instantaneous quenching from a high temperature to low temperatures. 
The equilibrium relaxation time increases in a super-Arrhenius manner with decreasing temperature. 
Despite this, the crystallization time shows an Arrhenius temperature dependence at very low temperature, and as a result, the crystallization becomes much faster than the equilibrium relaxation. 
This is contrary to the conventional expression of the crystallization time Eq.~(\ref{eq:convention}) based on the CNT. 
Furthermore, the estimated energy barrier for the crystallization is surprisingly small compared to those for the equilibrium dynamics.

To discuss this result, we first performed an energy topographic analysis. 
This analysis reveals that the crystallization process can be divided into an early process, in which the system falls into the nearest IS, and a late process, in which transitions between ISs occur. 
We show that the temperature dependence of the crystallization time is dominated by the late process. 
Second, we analyze the aging dynamics with and without the crystallization process. 
The aging dynamics consist of short-time relaxation, in which only the fastest particles undergo rearrangements and the correlation function decays exponentially, and the long-time relaxation, in which the majority of particles undergo rearrangements and the correlation function shows a long-time tail. 
In the case of aging without crystallization, the characteristic time scales of these dynamics are far from the crystallization time. 
Instead, in aging with crystallization, the relaxation is accelerated by crystallization, and the resulting relaxation time is quantitatively equivalent to the crystallization time. 
This means that a coupling between aging and crystallization is the key for understanding the rapid crystallization in deeply supercooled liquids or glasses. 

Previously, the crystallization of hard spheres at high density was studied, and various peculiar properties of crystallization dynamics have been reported, including the intermittent and avalanche-like crystallization dynamics and similarities between crystallization and aging dynamics ~\cite{de2017brownian,pusey2009hard,rosales2016avalanche,sanz2011crystallization,sanz2014avalanches,yanagishima2017common,zaccarelli2009crystallization}. 
Our work is partly consistent with these reports. 
Intermittent crystallization takes place not only in hard spheres but also soft spheres, suggesting that these peculiar crystallization dynamics are universal in deeply supercooled liquids or glasses.  
Moreover, our work reveals several new aspects of this crystallization. 
Most importantly, we could discuss the role of temperature in the crystallization dynamics and find the Arrhenius temperature dependence of the crystallization time with a very small energy barrier. 
Our findings would be relevant to understanding the crystallization of atomic and molecular glasses, where the temperature is the key control parameter. 
Since the model studied in this paper is a glass made by instantaneous quench from high temperature, it can be seen as a simplified model of molecular glasses made by vapor deposition at very low temperature, where the cooling rate is extremely large~\cite{tatsumi2012thermodynamic}. 
Our results suggest that the crystallization in such systems proceeds very rapidly with an Arrhenius temperature dependence even if they are fragile glass formers. 

In this paper, we focus solely on the instantaneous quench of the monodisperse system. 
It is interesting to extend our work to different preparation protocols and various types of glass formers. 
Such studies would enable us to understand the role of the maturity of glasses on the crystallization dynamics and the glass-forming ability at very low temperatures. 
It is also interesting to analyze the intermittency of the crystallization dynamics. 
In the case of the yielding of sheared glasses, the intermittency and avalanche formation have been studied in detail~\cite{oyama2021unified, maloney2006amorphous, lin2014scaling}, but a similar analysis seems to be lacking for systems driven by small thermal agitation. 
Such studies would provide more microscopic understandings of the crystallization and aging at very low temperatures. 

\begin{acknowledgments}
We thank Koji Hukushima for useful discussions.
This work is supported by JSPS KAKENHI (Grant Numbers 18H05225, 20H00128, 20H01868, 22K03543). 
\end{acknowledgments}

\bibliography{apssamp}

\begin{thebibliography}{41}%
\makeatletter
\providecommand \@ifxundefined [1]{%
 \@ifx{#1\undefined}
}%
\providecommand \@ifnum [1]{%
 \ifnum #1\expandafter \@firstoftwo
 \else \expandafter \@secondoftwo
 \fi
}%
\providecommand \@ifx [1]{%
 \ifx #1\expandafter \@firstoftwo
 \else \expandafter \@secondoftwo
 \fi
}%
\providecommand \natexlab [1]{#1}%
\providecommand \enquote  [1]{``#1''}%
\providecommand \bibnamefont  [1]{#1}%
\providecommand \bibfnamefont [1]{#1}%
\providecommand \citenamefont [1]{#1}%
\providecommand \href@noop [0]{\@secondoftwo}%
\providecommand \href [0]{\begingroup \@sanitize@url \@href}%
\providecommand \@href[1]{\@@startlink{#1}\@@href}%
\providecommand \@@href[1]{\endgroup#1\@@endlink}%
\providecommand \@sanitize@url [0]{\catcode `\\12\catcode `\$12\catcode
  `\&12\catcode `\#12\catcode `\^12\catcode `\_12\catcode `\%12\relax}%
\providecommand \@@startlink[1]{}%
\providecommand \@@endlink[0]{}%
\providecommand \url  [0]{\begingroup\@sanitize@url \@url }%
\providecommand \@url [1]{\endgroup\@href {#1}{\urlprefix }}%
\providecommand \urlprefix  [0]{URL }%
\providecommand \Eprint [0]{\href }%
\providecommand \doibase [0]{https://doi.org/}%
\providecommand \selectlanguage [0]{\@gobble}%
\providecommand \bibinfo  [0]{\@secondoftwo}%
\providecommand \bibfield  [0]{\@secondoftwo}%
\providecommand \translation [1]{[#1]}%
\providecommand \BibitemOpen [0]{}%
\providecommand \bibitemStop [0]{}%
\providecommand \bibitemNoStop [0]{.\EOS\space}%
\providecommand \EOS [0]{\spacefactor3000\relax}%
\providecommand \BibitemShut  [1]{\csname bibitem#1\endcsname}%
\let\auto@bib@innerbib\@empty
\bibitem [{\citenamefont {Debenedetti}(1997)}]{debenedetti1997metastable}%
  \BibitemOpen
  \bibfield  {author} {\bibinfo {author} {\bibfnamefont {P.~G.}\ \bibnamefont
  {Debenedetti}},\ }\href {https://doi.org/doi:10.1515/9780691213941} {\emph
  {\bibinfo {title} {Metastable Liquids}}}\ (\bibinfo  {publisher} {Princeton
  University Press},\ \bibinfo {address} {Princeton},\ \bibinfo {year}
  {1997})\BibitemShut {NoStop}%
\bibitem [{\citenamefont {Onuki}(2002)}]{onuki2002phase}%
  \BibitemOpen
  \bibfield  {author} {\bibinfo {author} {\bibfnamefont {A.}~\bibnamefont
  {Onuki}},\ }\href {https://books.google.co.jp/books?id=DA6AAVfv11sC} {\emph
  {\bibinfo {title} {Phase Transition Dynamics}}}\ (\bibinfo  {publisher}
  {Cambridge University Press},\ \bibinfo {year} {2002})\BibitemShut {NoStop}%
\bibitem [{\citenamefont {Binder}(1987)}]{Binder_1987}%
  \BibitemOpen
  \bibfield  {author} {\bibinfo {author} {\bibfnamefont {K.}~\bibnamefont
  {Binder}},\ }\bibfield  {title} {\bibinfo {title} {Theory of first-order
  phase transitions},\ }\href {https://doi.org/10.1088/0034-4885/50/7/001}
  {\bibfield  {journal} {\bibinfo  {journal} {Reports on Progress in Physics}\
  }\textbf {\bibinfo {volume} {50}},\ \bibinfo {pages} {783} (\bibinfo {year}
  {1987})}\BibitemShut {NoStop}%
\bibitem [{\citenamefont {Stauffer}\ \emph {et~al.}(1982)\citenamefont
  {Stauffer}, \citenamefont {Coniglio},\ and\ \citenamefont
  {Heermann}}]{Stauffer_1982}%
  \BibitemOpen
  \bibfield  {author} {\bibinfo {author} {\bibfnamefont {D.}~\bibnamefont
  {Stauffer}}, \bibinfo {author} {\bibfnamefont {A.}~\bibnamefont {Coniglio}},\
  and\ \bibinfo {author} {\bibfnamefont {D.~W.}\ \bibnamefont {Heermann}},\
  }\bibfield  {title} {\bibinfo {title} {Monte carlo experiment for nucleation
  rate in the three-dimensional ising model},\ }\href
  {https://doi.org/10.1103/PhysRevLett.49.1299} {\bibfield  {journal} {\bibinfo
   {journal} {Phys. Rev. Lett.}\ }\textbf {\bibinfo {volume} {49}},\ \bibinfo
  {pages} {1299} (\bibinfo {year} {1982})}\BibitemShut {NoStop}%
\bibitem [{\citenamefont {Shneidman}\ \emph {et~al.}(1999)\citenamefont
  {Shneidman}, \citenamefont {Jackson},\ and\ \citenamefont
  {Beatty}}]{Shneidman_1999}%
  \BibitemOpen
  \bibfield  {author} {\bibinfo {author} {\bibfnamefont {V.~A.}\ \bibnamefont
  {Shneidman}}, \bibinfo {author} {\bibfnamefont {K.~A.}\ \bibnamefont
  {Jackson}},\ and\ \bibinfo {author} {\bibfnamefont {K.~M.}\ \bibnamefont
  {Beatty}},\ }\bibfield  {title} {\bibinfo {title} {On the applicability of
  the classical nucleation theory in an ising system},\ }\href
  {https://doi.org/10.1063/1.479985} {\bibfield  {journal} {\bibinfo  {journal}
  {The Journal of Chemical Physics}\ }\textbf {\bibinfo {volume} {111}},\
  \bibinfo {pages} {6932} (\bibinfo {year} {1999})}\BibitemShut {NoStop}%
\bibitem [{\citenamefont {Ryu}\ and\ \citenamefont {Cai}(2010)}]{Ryu_2010}%
  \BibitemOpen
  \bibfield  {author} {\bibinfo {author} {\bibfnamefont {S.}~\bibnamefont
  {Ryu}}\ and\ \bibinfo {author} {\bibfnamefont {W.}~\bibnamefont {Cai}},\
  }\bibfield  {title} {\bibinfo {title} {Validity of classical nucleation
  theory for ising models},\ }\href
  {https://doi.org/10.1103/PhysRevE.81.030601} {\bibfield  {journal} {\bibinfo
  {journal} {Phys. Rev. E}\ }\textbf {\bibinfo {volume} {81}},\ \bibinfo
  {pages} {030601} (\bibinfo {year} {2010})}\BibitemShut {NoStop}%
\bibitem [{\citenamefont {Lundrigan}\ and\ \citenamefont
  {Saika-Voivod}(2009)}]{lundrigan2009test}%
  \BibitemOpen
  \bibfield  {author} {\bibinfo {author} {\bibfnamefont {S.~E.}\ \bibnamefont
  {Lundrigan}}\ and\ \bibinfo {author} {\bibfnamefont {I.}~\bibnamefont
  {Saika-Voivod}},\ }\bibfield  {title} {\bibinfo {title} {Test of classical
  nucleation theory and mean first-passage time formalism on crystallization in
  the lennard-jones liquid},\ }\href@noop {} {\bibfield  {journal} {\bibinfo
  {journal} {The Journal of Chemical Physics}\ }\textbf {\bibinfo {volume}
  {131}},\ \bibinfo {pages} {104503} (\bibinfo {year} {2009})}\BibitemShut
  {NoStop}%
\bibitem [{\citenamefont {Chkonia}\ \emph {et~al.}(2009)\citenamefont
  {Chkonia}, \citenamefont {W{\"o}lk}, \citenamefont {Strey}, \citenamefont
  {Wedekind},\ and\ \citenamefont {Reguera}}]{chkonia2009evaluating}%
  \BibitemOpen
  \bibfield  {author} {\bibinfo {author} {\bibfnamefont {G.}~\bibnamefont
  {Chkonia}}, \bibinfo {author} {\bibfnamefont {J.}~\bibnamefont {W{\"o}lk}},
  \bibinfo {author} {\bibfnamefont {R.}~\bibnamefont {Strey}}, \bibinfo
  {author} {\bibfnamefont {J.}~\bibnamefont {Wedekind}},\ and\ \bibinfo
  {author} {\bibfnamefont {D.}~\bibnamefont {Reguera}},\ }\bibfield  {title}
  {\bibinfo {title} {Evaluating nucleation rates in direct simulations},\
  }\href@noop {} {\bibfield  {journal} {\bibinfo  {journal} {The Journal of
  chemical physics}\ }\textbf {\bibinfo {volume} {130}},\ \bibinfo {pages}
  {064505} (\bibinfo {year} {2009})}\BibitemShut {NoStop}%
\bibitem [{\citenamefont {Auer}\ and\ \citenamefont
  {Frenkel}(2001)}]{Auer_2001}%
  \BibitemOpen
  \bibfield  {author} {\bibinfo {author} {\bibfnamefont {S.}~\bibnamefont
  {Auer}}\ and\ \bibinfo {author} {\bibfnamefont {D.}~\bibnamefont {Frenkel}},\
  }\bibfield  {title} {\bibinfo {title} {Prediction of absolute
  crystal-nucleation rate in hard-sphere colloids},\ }\href
  {https://doi.org/10.1038/35059035} {\bibfield  {journal} {\bibinfo  {journal}
  {Nature}\ }\textbf {\bibinfo {volume} {409}},\ \bibinfo {pages} {1020}
  (\bibinfo {year} {2001})}\BibitemShut {NoStop}%
\bibitem [{\citenamefont {Kawasaki}\ and\ \citenamefont
  {Tanaka}(2010)}]{Kawasaki_2010}%
  \BibitemOpen
  \bibfield  {author} {\bibinfo {author} {\bibfnamefont {T.}~\bibnamefont
  {Kawasaki}}\ and\ \bibinfo {author} {\bibfnamefont {H.}~\bibnamefont
  {Tanaka}},\ }\bibfield  {title} {\bibinfo {title} {Formation of a crystal
  nucleus from liquid},\ }\href {https://doi.org/10.1073/pnas.1001040107}
  {\bibfield  {journal} {\bibinfo  {journal} {Proceedings of the National
  Academy of Sciences}\ }\textbf {\bibinfo {volume} {107}},\ \bibinfo {pages}
  {14036} (\bibinfo {year} {2010})}\BibitemShut {NoStop}%
\bibitem [{\citenamefont {Cavagna}(2009)}]{cavagna2009supercooled}%
  \BibitemOpen
  \bibfield  {author} {\bibinfo {author} {\bibfnamefont {A.}~\bibnamefont
  {Cavagna}},\ }\bibfield  {title} {\bibinfo {title} {Supercooled liquids for
  pedestrians},\ }\href@noop {} {\bibfield  {journal} {\bibinfo  {journal}
  {Physics Reports}\ }\textbf {\bibinfo {volume} {476}},\ \bibinfo {pages} {51}
  (\bibinfo {year} {2009})}\BibitemShut {NoStop}%
\bibitem [{Note1()}]{Note1}%
  \BibitemOpen
  \bibinfo {note} {Although the inverse of the diffusion constant $D^{-1}$ is a
  better approximation of $\tau _t$~\cite {Tanaka_2003,Kawasaki_2010}, we omit
  the difference between $D^{-1}$ and $\tau _{\alpha }$. These two quantities
  show slightly different temperature dependences due to the Stokes-Einstein
  violation~\cite {cavagna2009supercooled}, but their difference is still minor
  for our later discussion.}\BibitemShut {Stop}%
\bibitem [{\citenamefont {Masuhr}\ \emph {et~al.}(1999)\citenamefont {Masuhr},
  \citenamefont {Waniuk}, \citenamefont {Busch},\ and\ \citenamefont
  {Johnson}}]{masuhr1999time}%
  \BibitemOpen
  \bibfield  {author} {\bibinfo {author} {\bibfnamefont {A.}~\bibnamefont
  {Masuhr}}, \bibinfo {author} {\bibfnamefont {T.}~\bibnamefont {Waniuk}},
  \bibinfo {author} {\bibfnamefont {R.}~\bibnamefont {Busch}},\ and\ \bibinfo
  {author} {\bibfnamefont {W.}~\bibnamefont {Johnson}},\ }\bibfield  {title}
  {\bibinfo {title} {Time scales for viscous flow, atomic transport, and
  crystallization in the liquid and supercooled liquid states of zr 41.2 ti
  13.8 cu 12.5 ni 10.0 be 22.5},\ }\href@noop {} {\bibfield  {journal}
  {\bibinfo  {journal} {Physical Review Letters}\ }\textbf {\bibinfo {volume}
  {82}},\ \bibinfo {pages} {2290} (\bibinfo {year} {1999})}\BibitemShut
  {NoStop}%
\bibitem [{\citenamefont {Kim}\ \emph {et~al.}(1996)\citenamefont {Kim},
  \citenamefont {Busch}, \citenamefont {Johnson}, \citenamefont {Rulison},\
  and\ \citenamefont {Rhim}}]{kim1996experimental}%
  \BibitemOpen
  \bibfield  {author} {\bibinfo {author} {\bibfnamefont {Y.}~\bibnamefont
  {Kim}}, \bibinfo {author} {\bibfnamefont {R.}~\bibnamefont {Busch}}, \bibinfo
  {author} {\bibfnamefont {W.}~\bibnamefont {Johnson}}, \bibinfo {author}
  {\bibfnamefont {A.}~\bibnamefont {Rulison}},\ and\ \bibinfo {author}
  {\bibfnamefont {W.}~\bibnamefont {Rhim}},\ }\bibfield  {title} {\bibinfo
  {title} {Experimental determination of a time--temperature-transformation
  diagram of the undercooled zr41. 2ti13. 8cu12. 5ni10. 0be22. 5 alloy using
  the containerless electrostatic levitation processing technique},\
  }\href@noop {} {\bibfield  {journal} {\bibinfo  {journal} {Applied Physics
  Letters}\ }\textbf {\bibinfo {volume} {68}},\ \bibinfo {pages} {1057}
  (\bibinfo {year} {1996})}\BibitemShut {NoStop}%
\bibitem [{\citenamefont {Schroers}\ \emph {et~al.}(2001)\citenamefont
  {Schroers}, \citenamefont {Wu}, \citenamefont {Busch},\ and\ \citenamefont
  {Johnson}}]{schroers2001transition}%
  \BibitemOpen
  \bibfield  {author} {\bibinfo {author} {\bibfnamefont {J.}~\bibnamefont
  {Schroers}}, \bibinfo {author} {\bibfnamefont {Y.}~\bibnamefont {Wu}},
  \bibinfo {author} {\bibfnamefont {R.}~\bibnamefont {Busch}},\ and\ \bibinfo
  {author} {\bibfnamefont {W.}~\bibnamefont {Johnson}},\ }\bibfield  {title}
  {\bibinfo {title} {Transition from nucleation controlled to growth controlled
  crystallization in pd43ni10cu27p20 melts},\ }\href@noop {} {\bibfield
  {journal} {\bibinfo  {journal} {Acta Materialia}\ }\textbf {\bibinfo {volume}
  {49}},\ \bibinfo {pages} {2773} (\bibinfo {year} {2001})}\BibitemShut
  {NoStop}%
\bibitem [{\citenamefont {Levchenko}\ \emph {et~al.}(2011)\citenamefont
  {Levchenko}, \citenamefont {Evteev}, \citenamefont {Belova},\ and\
  \citenamefont {Murch}}]{levchenko2011molecular}%
  \BibitemOpen
  \bibfield  {author} {\bibinfo {author} {\bibfnamefont {E.~V.}\ \bibnamefont
  {Levchenko}}, \bibinfo {author} {\bibfnamefont {A.~V.}\ \bibnamefont
  {Evteev}}, \bibinfo {author} {\bibfnamefont {I.~V.}\ \bibnamefont {Belova}},\
  and\ \bibinfo {author} {\bibfnamefont {G.~E.}\ \bibnamefont {Murch}},\
  }\bibfield  {title} {\bibinfo {title} {Molecular dynamics determination of
  the time--temperature--transformation diagram for crystallization of an
  undercooled liquid ni50al50 alloy},\ }\href@noop {} {\bibfield  {journal}
  {\bibinfo  {journal} {Acta materialia}\ }\textbf {\bibinfo {volume} {59}},\
  \bibinfo {pages} {6412} (\bibinfo {year} {2011})}\BibitemShut {NoStop}%
\bibitem [{\citenamefont {Moore}\ and\ \citenamefont
  {Molinero}(2011)}]{moore2011structural}%
  \BibitemOpen
  \bibfield  {author} {\bibinfo {author} {\bibfnamefont {E.~B.}\ \bibnamefont
  {Moore}}\ and\ \bibinfo {author} {\bibfnamefont {V.}~\bibnamefont
  {Molinero}},\ }\bibfield  {title} {\bibinfo {title} {Structural
  transformation in supercooled water controls the crystallization rate of
  ice},\ }\href@noop {} {\bibfield  {journal} {\bibinfo  {journal} {Nature}\
  }\textbf {\bibinfo {volume} {479}},\ \bibinfo {pages} {506} (\bibinfo {year}
  {2011})}\BibitemShut {NoStop}%
\bibitem [{\citenamefont {Zaccarelli}\ \emph {et~al.}(2009)\citenamefont
  {Zaccarelli}, \citenamefont {Valeriani}, \citenamefont {Sanz}, \citenamefont
  {Poon}, \citenamefont {Cates},\ and\ \citenamefont
  {Pusey}}]{zaccarelli2009crystallization}%
  \BibitemOpen
  \bibfield  {author} {\bibinfo {author} {\bibfnamefont {E.}~\bibnamefont
  {Zaccarelli}}, \bibinfo {author} {\bibfnamefont {C.}~\bibnamefont
  {Valeriani}}, \bibinfo {author} {\bibfnamefont {E.}~\bibnamefont {Sanz}},
  \bibinfo {author} {\bibfnamefont {W.}~\bibnamefont {Poon}}, \bibinfo {author}
  {\bibfnamefont {M.}~\bibnamefont {Cates}},\ and\ \bibinfo {author}
  {\bibfnamefont {P.}~\bibnamefont {Pusey}},\ }\bibfield  {title} {\bibinfo
  {title} {Crystallization of hard-sphere glasses},\ }\href@noop {} {\bibfield
  {journal} {\bibinfo  {journal} {Physical review letters}\ }\textbf {\bibinfo
  {volume} {103}},\ \bibinfo {pages} {135704} (\bibinfo {year}
  {2009})}\BibitemShut {NoStop}%
\bibitem [{\citenamefont {Sanz}\ \emph {et~al.}(2011)\citenamefont {Sanz},
  \citenamefont {Valeriani}, \citenamefont {Zaccarelli}, \citenamefont {Poon},
  \citenamefont {Pusey},\ and\ \citenamefont
  {Cates}}]{sanz2011crystallization}%
  \BibitemOpen
  \bibfield  {author} {\bibinfo {author} {\bibfnamefont {E.}~\bibnamefont
  {Sanz}}, \bibinfo {author} {\bibfnamefont {C.}~\bibnamefont {Valeriani}},
  \bibinfo {author} {\bibfnamefont {E.}~\bibnamefont {Zaccarelli}}, \bibinfo
  {author} {\bibfnamefont {W.~C.}\ \bibnamefont {Poon}}, \bibinfo {author}
  {\bibfnamefont {P.~N.}\ \bibnamefont {Pusey}},\ and\ \bibinfo {author}
  {\bibfnamefont {M.~E.}\ \bibnamefont {Cates}},\ }\bibfield  {title} {\bibinfo
  {title} {Crystallization mechanism of hard sphere glasses},\ }\href@noop {}
  {\bibfield  {journal} {\bibinfo  {journal} {Physical Review Letters}\
  }\textbf {\bibinfo {volume} {106}},\ \bibinfo {pages} {215701} (\bibinfo
  {year} {2011})}\BibitemShut {NoStop}%
\bibitem [{\citenamefont {Sanz}\ \emph {et~al.}(2014)\citenamefont {Sanz},
  \citenamefont {Valeriani}, \citenamefont {Zaccarelli}, \citenamefont {Poon},
  \citenamefont {Cates},\ and\ \citenamefont {Pusey}}]{sanz2014avalanches}%
  \BibitemOpen
  \bibfield  {author} {\bibinfo {author} {\bibfnamefont {E.}~\bibnamefont
  {Sanz}}, \bibinfo {author} {\bibfnamefont {C.}~\bibnamefont {Valeriani}},
  \bibinfo {author} {\bibfnamefont {E.}~\bibnamefont {Zaccarelli}}, \bibinfo
  {author} {\bibfnamefont {W.~C.}\ \bibnamefont {Poon}}, \bibinfo {author}
  {\bibfnamefont {M.~E.}\ \bibnamefont {Cates}},\ and\ \bibinfo {author}
  {\bibfnamefont {P.~N.}\ \bibnamefont {Pusey}},\ }\bibfield  {title} {\bibinfo
  {title} {Avalanches mediate crystallization in a hard-sphere glass},\
  }\href@noop {} {\bibfield  {journal} {\bibinfo  {journal} {Proceedings of the
  National Academy of Sciences}\ }\textbf {\bibinfo {volume} {111}},\ \bibinfo
  {pages} {75} (\bibinfo {year} {2014})}\BibitemShut {NoStop}%
\bibitem [{\citenamefont {de~Hijes}\ \emph {et~al.}(2017)\citenamefont
  {de~Hijes}, \citenamefont {Rosales-Pelaez}, \citenamefont {Valeriani},
  \citenamefont {Pusey},\ and\ \citenamefont {Sanz}}]{de2017brownian}%
  \BibitemOpen
  \bibfield  {author} {\bibinfo {author} {\bibfnamefont {P.~M.}\ \bibnamefont
  {de~Hijes}}, \bibinfo {author} {\bibfnamefont {P.}~\bibnamefont
  {Rosales-Pelaez}}, \bibinfo {author} {\bibfnamefont {C.}~\bibnamefont
  {Valeriani}}, \bibinfo {author} {\bibfnamefont {P.~N.}\ \bibnamefont
  {Pusey}},\ and\ \bibinfo {author} {\bibfnamefont {E.}~\bibnamefont {Sanz}},\
  }\bibfield  {title} {\bibinfo {title} {Brownian versus newtonian
  devitrification of hard-sphere glasses},\ }\href@noop {} {\bibfield
  {journal} {\bibinfo  {journal} {Physical Review E}\ }\textbf {\bibinfo
  {volume} {96}},\ \bibinfo {pages} {020602} (\bibinfo {year}
  {2017})}\BibitemShut {NoStop}%
\bibitem [{\citenamefont {Rosales-Pelaez}\ \emph {et~al.}(2016)\citenamefont
  {Rosales-Pelaez}, \citenamefont {de~Hijes}, \citenamefont {Sanz},\ and\
  \citenamefont {Valeriani}}]{rosales2016avalanche}%
  \BibitemOpen
  \bibfield  {author} {\bibinfo {author} {\bibfnamefont {P.}~\bibnamefont
  {Rosales-Pelaez}}, \bibinfo {author} {\bibfnamefont {P.~M.}\ \bibnamefont
  {de~Hijes}}, \bibinfo {author} {\bibfnamefont {E.}~\bibnamefont {Sanz}},\
  and\ \bibinfo {author} {\bibfnamefont {C.}~\bibnamefont {Valeriani}},\
  }\bibfield  {title} {\bibinfo {title} {Avalanche mediated devitrification in
  a glass of pseudo hard-spheres},\ }\href@noop {} {\bibfield  {journal}
  {\bibinfo  {journal} {Journal of Statistical Mechanics: Theory and
  Experiment}\ }\textbf {\bibinfo {volume} {2016}},\ \bibinfo {pages} {094005}
  (\bibinfo {year} {2016})}\BibitemShut {NoStop}%
\bibitem [{\citenamefont {Yanagishima}\ \emph {et~al.}(2017)\citenamefont
  {Yanagishima}, \citenamefont {Russo},\ and\ \citenamefont
  {Tanaka}}]{yanagishima2017common}%
  \BibitemOpen
  \bibfield  {author} {\bibinfo {author} {\bibfnamefont {T.}~\bibnamefont
  {Yanagishima}}, \bibinfo {author} {\bibfnamefont {J.}~\bibnamefont {Russo}},\
  and\ \bibinfo {author} {\bibfnamefont {H.}~\bibnamefont {Tanaka}},\
  }\bibfield  {title} {\bibinfo {title} {Common mechanism of thermodynamic and
  mechanical origin for ageing and crystallization of glasses},\ }\href@noop {}
  {\bibfield  {journal} {\bibinfo  {journal} {Nature communications}\ }\textbf
  {\bibinfo {volume} {8}},\ \bibinfo {pages} {1} (\bibinfo {year}
  {2017})}\BibitemShut {NoStop}%
\bibitem [{\citenamefont {Ganapathi}\ \emph {et~al.}(2021)\citenamefont
  {Ganapathi}, \citenamefont {Chakrabarti}, \citenamefont {Sood},\ and\
  \citenamefont {Ganapathy}}]{ganapathi2021structure}%
  \BibitemOpen
  \bibfield  {author} {\bibinfo {author} {\bibfnamefont {D.}~\bibnamefont
  {Ganapathi}}, \bibinfo {author} {\bibfnamefont {D.}~\bibnamefont
  {Chakrabarti}}, \bibinfo {author} {\bibfnamefont {A.}~\bibnamefont {Sood}},\
  and\ \bibinfo {author} {\bibfnamefont {R.}~\bibnamefont {Ganapathy}},\
  }\bibfield  {title} {\bibinfo {title} {Structure determines where
  crystallization occurs in a soft colloidal glass},\ }\href@noop {} {\bibfield
   {journal} {\bibinfo  {journal} {Nature Physics}\ }\textbf {\bibinfo {volume}
  {17}},\ \bibinfo {pages} {114} (\bibinfo {year} {2021})}\BibitemShut
  {NoStop}%
\bibitem [{\citenamefont {Zaccarelli}\ \emph {et~al.}(2015)\citenamefont
  {Zaccarelli}, \citenamefont {Liddle},\ and\ \citenamefont
  {Poon}}]{zaccarelli2015polydispersity}%
  \BibitemOpen
  \bibfield  {author} {\bibinfo {author} {\bibfnamefont {E.}~\bibnamefont
  {Zaccarelli}}, \bibinfo {author} {\bibfnamefont {S.~M.}\ \bibnamefont
  {Liddle}},\ and\ \bibinfo {author} {\bibfnamefont {W.~C.}\ \bibnamefont
  {Poon}},\ }\bibfield  {title} {\bibinfo {title} {On polydispersity and the
  hard sphere glass transition},\ }\href@noop {} {\bibfield  {journal}
  {\bibinfo  {journal} {Soft Matter}\ }\textbf {\bibinfo {volume} {11}},\
  \bibinfo {pages} {324} (\bibinfo {year} {2015})}\BibitemShut {NoStop}%
\bibitem [{Note2()}]{Note2}%
  \BibitemOpen
  \bibinfo {note} {We calculated the free energy of the equilibrium liquid and
  crystal with Monte Carlo simulations. The calculations were conducted
  according to the method described in Ref.~\cite {prestipino2005phase}, which
  we do not describe in detail.}\BibitemShut {Stop}%
\bibitem [{\citenamefont {Pusey}\ \emph {et~al.}(2009)\citenamefont {Pusey},
  \citenamefont {Zaccarelli}, \citenamefont {Valeriani}, \citenamefont {Sanz},
  \citenamefont {Poon},\ and\ \citenamefont {Cates}}]{pusey2009hard}%
  \BibitemOpen
  \bibfield  {author} {\bibinfo {author} {\bibfnamefont {P.}~\bibnamefont
  {Pusey}}, \bibinfo {author} {\bibfnamefont {E.}~\bibnamefont {Zaccarelli}},
  \bibinfo {author} {\bibfnamefont {C.}~\bibnamefont {Valeriani}}, \bibinfo
  {author} {\bibfnamefont {E.}~\bibnamefont {Sanz}}, \bibinfo {author}
  {\bibfnamefont {W.~C.}\ \bibnamefont {Poon}},\ and\ \bibinfo {author}
  {\bibfnamefont {M.~E.}\ \bibnamefont {Cates}},\ }\bibfield  {title} {\bibinfo
  {title} {Hard spheres: crystallization and glass formation},\ }\href@noop {}
  {\bibfield  {journal} {\bibinfo  {journal} {Philosophical Transactions of the
  Royal Society A: Mathematical, Physical and Engineering Sciences}\ }\textbf
  {\bibinfo {volume} {367}},\ \bibinfo {pages} {4993} (\bibinfo {year}
  {2009})}\BibitemShut {NoStop}%
\bibitem [{\citenamefont {ten Wolde}\ \emph {et~al.}(1996)\citenamefont {ten
  Wolde}, \citenamefont {Ruiz-Montero},\ and\ \citenamefont
  {Frenkel}}]{wolde1996simulation}%
  \BibitemOpen
  \bibfield  {author} {\bibinfo {author} {\bibfnamefont {P.-R.}\ \bibnamefont
  {ten Wolde}}, \bibinfo {author} {\bibfnamefont {M.~J.}\ \bibnamefont
  {Ruiz-Montero}},\ and\ \bibinfo {author} {\bibfnamefont {D.}~\bibnamefont
  {Frenkel}},\ }\bibfield  {title} {\bibinfo {title} {Simulation of homogeneous
  crystal nucleation close to coexistence},\ }\href
  {https://doi.org/10.1039/FD9960400093} {\bibfield  {journal} {\bibinfo
  {journal} {Faraday Discuss.}\ }\textbf {\bibinfo {volume} {104}},\ \bibinfo
  {pages} {93} (\bibinfo {year} {1996})}\BibitemShut {NoStop}%
\bibitem [{\citenamefont {Kawasaki}\ \emph {et~al.}(2007)\citenamefont
  {Kawasaki}, \citenamefont {Araki},\ and\ \citenamefont
  {Tanaka}}]{kawasaki2007correlation}%
  \BibitemOpen
  \bibfield  {author} {\bibinfo {author} {\bibfnamefont {T.}~\bibnamefont
  {Kawasaki}}, \bibinfo {author} {\bibfnamefont {T.}~\bibnamefont {Araki}},\
  and\ \bibinfo {author} {\bibfnamefont {H.}~\bibnamefont {Tanaka}},\
  }\bibfield  {title} {\bibinfo {title} {Correlation between dynamic
  heterogeneity and medium-range order in two-dimensional glass-forming
  liquids},\ }\href@noop {} {\bibfield  {journal} {\bibinfo  {journal}
  {Physical review letters}\ }\textbf {\bibinfo {volume} {99}},\ \bibinfo
  {pages} {215701} (\bibinfo {year} {2007})}\BibitemShut {NoStop}%
\bibitem [{Note3()}]{Note3}%
  \BibitemOpen
  \bibinfo {note} {Note that the crystallization dynamics become slower in much
  smaller systems at lower temperatures. This slowing is a finite size effect
  due to the collisions of the crystalline regions which the periodic boundary
  condition facilitates. We confirmed that $N=16384$ is large enough for the
  focused temperature range.}\BibitemShut {Stop}%
\bibitem [{Note4()}]{Note4}%
  \BibitemOpen
  \bibinfo {note} {We confirmed that an increase or decrease in the threshold
  value 0.5 only causes a constant-fold change in the crystallization time,
  barely altering its temperature dependence.}\BibitemShut {Stop}%
\bibitem [{Note5()}]{Note5}%
  \BibitemOpen
  \bibinfo {note} {This fitting was done for the data in $0.3 < P(t) < 0.8$.
  Note that the error of the estimate of $\tau _{\protect \mathrm {lag}}$
  became larger when $\tau _{\protect \mathrm {exp}}$ became much larger than
  $\tau _{\protect \mathrm {lag}}$, and we could not estimate $\tau _{\protect
  \mathrm {lag}}$ at the lowest temperature.}\BibitemShut {Stop}%
\bibitem [{\citenamefont {Bitzek}\ \emph {et~al.}(2006)\citenamefont {Bitzek},
  \citenamefont {Koskinen}, \citenamefont {G{\"a}hler}, \citenamefont
  {Moseler},\ and\ \citenamefont {Gumbsch}}]{bitzek2006structural}%
  \BibitemOpen
  \bibfield  {author} {\bibinfo {author} {\bibfnamefont {E.}~\bibnamefont
  {Bitzek}}, \bibinfo {author} {\bibfnamefont {P.}~\bibnamefont {Koskinen}},
  \bibinfo {author} {\bibfnamefont {F.}~\bibnamefont {G{\"a}hler}}, \bibinfo
  {author} {\bibfnamefont {M.}~\bibnamefont {Moseler}},\ and\ \bibinfo {author}
  {\bibfnamefont {P.}~\bibnamefont {Gumbsch}},\ }\bibfield  {title} {\bibinfo
  {title} {Structural relaxation made simple},\ }\href@noop {} {\bibfield
  {journal} {\bibinfo  {journal} {Physical review letters}\ }\textbf {\bibinfo
  {volume} {97}},\ \bibinfo {pages} {170201} (\bibinfo {year}
  {2006})}\BibitemShut {NoStop}%
\bibitem [{\citenamefont {El~Masri}\ \emph {et~al.}(2010)\citenamefont
  {El~Masri}, \citenamefont {Berthier},\ and\ \citenamefont
  {Cipelletti}}]{el2010subdiffusion}%
  \BibitemOpen
  \bibfield  {author} {\bibinfo {author} {\bibfnamefont {D.}~\bibnamefont
  {El~Masri}}, \bibinfo {author} {\bibfnamefont {L.}~\bibnamefont {Berthier}},\
  and\ \bibinfo {author} {\bibfnamefont {L.}~\bibnamefont {Cipelletti}},\
  }\bibfield  {title} {\bibinfo {title} {Subdiffusion and intermittent dynamic
  fluctuations in the aging regime of concentrated hard spheres},\ }\href@noop
  {} {\bibfield  {journal} {\bibinfo  {journal} {Physical Review E}\ }\textbf
  {\bibinfo {volume} {82}},\ \bibinfo {pages} {031503} (\bibinfo {year}
  {2010})}\BibitemShut {NoStop}%
\bibitem [{Note6()}]{Note6}%
  \BibitemOpen
  \bibinfo {note} {For simplicity, we use the symbol $\tau _{\protect \mathrm
  {aging,mono}}$, although not only aging but also crystallization take place
  in this case.}\BibitemShut {Stop}%
\bibitem [{\citenamefont {Tatsumi}\ \emph {et~al.}(2012)\citenamefont
  {Tatsumi}, \citenamefont {Aso},\ and\ \citenamefont
  {Yamamuro}}]{tatsumi2012thermodynamic}%
  \BibitemOpen
  \bibfield  {author} {\bibinfo {author} {\bibfnamefont {S.}~\bibnamefont
  {Tatsumi}}, \bibinfo {author} {\bibfnamefont {S.}~\bibnamefont {Aso}},\ and\
  \bibinfo {author} {\bibfnamefont {O.}~\bibnamefont {Yamamuro}},\ }\bibfield
  {title} {\bibinfo {title} {Thermodynamic study of simple molecular glasses:
  universal features in their heat capacity and the size of the cooperatively
  rearranging regions},\ }\href@noop {} {\bibfield  {journal} {\bibinfo
  {journal} {Physical Review Letters}\ }\textbf {\bibinfo {volume} {109}},\
  \bibinfo {pages} {045701} (\bibinfo {year} {2012})}\BibitemShut {NoStop}%
\bibitem [{\citenamefont {Oyama}\ \emph {et~al.}(2021)\citenamefont {Oyama},
  \citenamefont {Mizuno},\ and\ \citenamefont {Ikeda}}]{oyama2021unified}%
  \BibitemOpen
  \bibfield  {author} {\bibinfo {author} {\bibfnamefont {N.}~\bibnamefont
  {Oyama}}, \bibinfo {author} {\bibfnamefont {H.}~\bibnamefont {Mizuno}},\ and\
  \bibinfo {author} {\bibfnamefont {A.}~\bibnamefont {Ikeda}},\ }\bibfield
  {title} {\bibinfo {title} {Unified view of avalanche criticality in sheared
  glasses},\ }\href@noop {} {\bibfield  {journal} {\bibinfo  {journal}
  {Physical Review E}\ }\textbf {\bibinfo {volume} {104}},\ \bibinfo {pages}
  {015002} (\bibinfo {year} {2021})}\BibitemShut {NoStop}%
\bibitem [{\citenamefont {Maloney}\ and\ \citenamefont
  {Lemaitre}(2006)}]{maloney2006amorphous}%
  \BibitemOpen
  \bibfield  {author} {\bibinfo {author} {\bibfnamefont {C.~E.}\ \bibnamefont
  {Maloney}}\ and\ \bibinfo {author} {\bibfnamefont {A.}~\bibnamefont
  {Lemaitre}},\ }\bibfield  {title} {\bibinfo {title} {Amorphous systems in
  athermal, quasistatic shear},\ }\href@noop {} {\bibfield  {journal} {\bibinfo
   {journal} {Physical Review E}\ }\textbf {\bibinfo {volume} {74}},\ \bibinfo
  {pages} {016118} (\bibinfo {year} {2006})}\BibitemShut {NoStop}%
\bibitem [{\citenamefont {Lin}\ \emph {et~al.}(2014)\citenamefont {Lin},
  \citenamefont {Lerner}, \citenamefont {Rosso},\ and\ \citenamefont
  {Wyart}}]{lin2014scaling}%
  \BibitemOpen
  \bibfield  {author} {\bibinfo {author} {\bibfnamefont {J.}~\bibnamefont
  {Lin}}, \bibinfo {author} {\bibfnamefont {E.}~\bibnamefont {Lerner}},
  \bibinfo {author} {\bibfnamefont {A.}~\bibnamefont {Rosso}},\ and\ \bibinfo
  {author} {\bibfnamefont {M.}~\bibnamefont {Wyart}},\ }\bibfield  {title}
  {\bibinfo {title} {Scaling description of the yielding transition in soft
  amorphous solids at zero temperature},\ }\href@noop {} {\bibfield  {journal}
  {\bibinfo  {journal} {Proceedings of the National Academy of Sciences}\
  }\textbf {\bibinfo {volume} {111}},\ \bibinfo {pages} {14382} (\bibinfo
  {year} {2014})}\BibitemShut {NoStop}%
\bibitem [{\citenamefont {Tanaka}(2003)}]{Tanaka_2003}%
  \BibitemOpen
  \bibfield  {author} {\bibinfo {author} {\bibfnamefont {H.}~\bibnamefont
  {Tanaka}},\ }\bibfield  {title} {\bibinfo {title} {Possible resolution of the
  kauzmann paradox in supercooled liquids},\ }\href
  {https://doi.org/10.1103/PhysRevE.68.011505} {\bibfield  {journal} {\bibinfo
  {journal} {Phys. Rev. E}\ }\textbf {\bibinfo {volume} {68}},\ \bibinfo
  {pages} {011505} (\bibinfo {year} {2003})}\BibitemShut {NoStop}%
\bibitem [{\citenamefont {Prestipino}\ \emph {et~al.}(2005)\citenamefont
  {Prestipino}, \citenamefont {Saija},\ and\ \citenamefont
  {Giaquinta}}]{prestipino2005phase}%
  \BibitemOpen
  \bibfield  {author} {\bibinfo {author} {\bibfnamefont {S.}~\bibnamefont
  {Prestipino}}, \bibinfo {author} {\bibfnamefont {F.}~\bibnamefont {Saija}},\
  and\ \bibinfo {author} {\bibfnamefont {P.~V.}\ \bibnamefont {Giaquinta}},\
  }\bibfield  {title} {\bibinfo {title} {Phase diagram of softly repulsive
  systems: The gaussian and inverse-power-law potentials},\ }\href@noop {}
  {\bibfield  {journal} {\bibinfo  {journal} {The Journal of chemical physics}\
  }\textbf {\bibinfo {volume} {123}},\ \bibinfo {pages} {144110} (\bibinfo
  {year} {2005})}\BibitemShut {NoStop}%
\end{thebibliography}%

\end{document}